\documentclass[aps,prd,onecolumn,groupedaddress,showpacs,nofootinbib,amssymb]{revtex4}

\usepackage[dvips]{graphicx}
\usepackage{amssymb}
\usepackage{amsmath}
\usepackage{amsfonts}
\usepackage{bm}
\usepackage{cancel}
\usepackage{comment}
\usepackage{color}
\usepackage{float}
\usepackage{hyperref} 

\newcommand\be{\begin{equation}}
\newcommand\ee{\end{equation}}

\allowdisplaybreaks[4]

\begin{document}

\tolerance=5000

\title{Mass-Gap Neutron Stars from Vector \texorpdfstring{$f(R)$}{f(R)} Gravity Inflationary Deformations}

\author{V.K. Oikonomou,$^{1,2}$}
\email{voikonomou@gapps.auth.gr,v.k.oikonomou1979@gmail.com}
\affiliation{$^{1)}$Department of Physics, Aristotle University of Thessaloniki, Thessaloniki 54124, Greece\\
$^{2)}$L.N. Gumilyov Eurasian National University - Astana,
010008, Kazakhstan}

\tolerance=5000

\begin{abstract}
The latest observations from the LIGO-Virgo indicated the
existence of mass-gap region astrophysical objects. This is a
rather sensational observation and there are two possibilities for
the nature of these mass-gap region astrophysical objects, these
are either small black holes that result from the mergers of
ordinary mass neutron stars, or these are heavy neutron stars. In
the line of research implied by the former possibility, in this
work we shall examine the implied neutron star phenomenology from
vector $f(R)$ gravity inflationary models. These theories are
basically scalar-tensor deformations of the Starobinsky
inflationary model. We shall present the essential features of
cosmologically viable and non-viable deformations of the
Starobinsky model, originating from vector $f(R)$ gravity
inflationary theories, and we indicate which models and for which
equations of state provide a viable neutron star phenomenology. We
solve the Tolman-Oppenheimer-Volkov equations using a robust
double shooting LSODA python based code, for the following
piecewise polytropic equations of state the WFF1, the SLy, the
APR, the MS1, the AP3, the AP4, the ENG, the MPA1 and the MS1b. We
confront the resulting phenomenology with several well known
neutron star constraints and we indicate which equation of state
and model fits the phenomenological constraints. A remarkable
feature, also known from other inflationary attractor models, is
that the MPA1 is the equation of state which is most nicely fitted
the constraints, for all the theoretical models used, and actually
the maximum mass for this equation of state is well inside the
mass-gap region. Another mentionable feature that stroked us with
surprise is the fact that even cosmologically non-viable
inflationary models produced a viable neutron star phenomenology,
which most likely has to be a model-dependent feature.
\end{abstract}


\maketitle

\section*{Introduction}

Recent astrophysical observations by LIGO-Virgo have pointed out
the existence of compact astrophysical objects with masses in the
mass-gap region, which is the range of masses $M\sim
2.5-5\,M_{\odot}$, see for example the event GW190814
\cite{Abbott:2020khf} or the more recent GW230529
\cite{LIGOScientific:2024elc}. Although the most possible
explanation for the identity of these objects is that these are
light black holes which result from the merging of two ordinary
mass neutron stars, there exists the sensational possibility that
these mass-gap region objects are neutron stars (NS)
\cite{Haensel:2007yy,Friedman:2013xza,Baym:2017whm,Lattimer:2004pg,Olmo:2019flu}.
Then the question emerging is, how are these NSs explained, on
what ground these are theoretically supported. This is not an easy
question to answer, since an explanation might be that General
Relativity (GR) in conjunction with a stiff equation of state
(EoS) might describe the existence of such heavy NSs. Or that
these NSs are described by some modification of GR. Thus there is
the ambiguity of heavy NSs, are these explained by a stiff EoS or
modified gravity? This is a difficult question to answer, however
we must have in mind that the EoS of NSs should be unique for all
the NS spanning a large mass range. Thus these stiff EoSs should
also be compatible with all the phenomenological constraints that
apply to NSs. To this end, modified gravity can accommodate large
NS masses rather naturally without relying to the stiffness of the
EoS. The modified gravity paradigm thus stands as a viable
explanation for mass-gap region NS. Noted that there is an upper
limit in the stiffness of the EoS of ordinary NSs, the causal
limit equation of state, which indicates that the maximum static
NS mass is 3 solar masses, within the context of GR
 \cite{Rhoades:1974fn,Kalogera:1996ci},
\begin{equation}\label{causalupperbound}
M_{max}^{CL}=3M_{\odot}\sqrt{\frac{5\times
10^{14}g/cm^{3}}{\rho_u}}\, ,
\end{equation}
with $\rho_u$ being the reference density that separates the
causal region and the low-density region. For the low-density
region, the EoS is known, and the corresponding pressure is
$P_u(\rho_u)$, and the exact causal EoS has the form,
\begin{equation}\label{causallimiteos}
P_{sn}(\rho)=P_{u}(\rho_u)+(\rho-\rho_u)c^2\, .
\end{equation}
Finally, for rotating NSs, the causal EoS maximum mass is,
\begin{equation}\label{causalrot}
M^{CL,rot}_{max}=3.89M_{\odot}\sqrt{\frac{5\times
10^{14}g/cm^{3}}{\rho_u}}\, .
\end{equation}
It is important to discuss at this point the perspective of
modified gravity. Highly spinning NSs are considered NSs that have
periods $P<3$ms, so basically millisecond pulsars. Thus if the NSs
has a larger period than 3ms, then one can safely approximate the
NS as a nearly static one since the stellar structure is not
significantly affected \cite{Haensel:2007yy}. Thus, if one
considers static NSs, the GR limit of the maximum mass is 3 solar
masses. It turns out that in most popular GR extensions, the 3
solar mass limit for the maximum static NSs is respected, see for
example \cite{Astashenok:2021peo}, and also
\cite{Odintsov:2023ypt,Oikonomou:2024yzj} for popular
scalar-tensor extensions of GR. The important thing here to note
that once static NSs are considered, maximum masses in the mass
gap region $2.5-3M_{\odot}$ cannot be described successfully with
GR, even when very stiff EoSs are used. Modified gravity can
actually describe such NSs without extreme fine tuning and in a
viable way for a large number of available EoSs. To our opinion,
finding NSs beyond 3 solar masses is unrealistic, even in the
context of modified gravity and such results should be carefully
interpreted. To date, there are objects in the mass-gap region
$2.5-5\,M_{\odot}$, but these are not confirmed to be NSs, and to
our opinion these are black holes probably emanating from the
merging of two NSs. Of course, if the compact objects in the
mass-gap region are confirmed to be NSs, there is the possibility
that these have a high spin thus can be described even in the
context of GR, if these are millisecond pulsars. We hope in the
near future nature will be kind to us and reveal its mysteries
regarding these issues.

The GW170817 event \cite{TheLIGOScientific:2017qsa} imposed some
strong constraints on the allowed EoS behavior for NSs. The event
GW170817 was very illuminating, since it was followed by a
kilonova thus confirming the merging of two NSs. The recent
mass-gap region related events
\cite{Abbott:2020khf,LIGOScientific:2024elc}, were not followed by
a kilonova, thus it is hard to speculate if heavy NSs were
involved. Certainly, a future observation of a kilonova event in a
merger of mass-gap region compact objects will verify if heavy NSs
exist in nature and if NSs can have masses in the range
$2.5-3\,M_{\odot}$, or even beyond 3 solar masses. Currently, the
highest mass NS ever observed is the low-spin pulsar known as
black widow pulsar PSR J0952-0607 with mass $M=2.35\pm 0.17$
\cite{Romani:2022jhd}, which is quite close to the mass-gap
region. Hopefully, if nature is kind with us and we are lucky
enough, the question whether modified gravity or some stiff EoS
can describe heavy NSs will be better understood in the next
decades. But still, there are a lot of issues to be better
understood, degeneracy between the EoS and the modified gravity
model, even degeneracies between different modified gravity models
and so on. In this work we shall adopt the modified gravity (for
reviews see \cite{reviews1,reviews2,reviews3,reviews4,reviews5})
explanation of heavy NSs, and we shall examine the phenomenology
of NSs produced by a class of vector $f(R)$ gravity inflationary
potentials \cite{Ozkan:2015iva}. Apparently, NS physics is in the
mainstream of modern theoretical physics research nowadays since
many different physics frameworks use NSs for their framework, for
example nuclear physics research
\cite{Lattimer:2012nd,Steiner:2011ft,Horowitz:2005zb,Watanabe:2000rj,Shen:1998gq,Xu:2009vi,Hebeler:2013nza,Mendoza-Temis:2014mja,Ho:2014pta,Kanakis-Pegios:2020kzp,Tsaloukidis:2022rus,Kanakis-Pegios:2023gvc},
high energy physics
\cite{Buschmann:2019pfp,Safdi:2018oeu,Hook:2018iia,Edwards:2020afl,Nurmi:2021xds},
modified gravity,
\cite{Astashenok:2020qds,Astashenok:2021peo,Capozziello:2015yza,Astashenok:2014nua,Astashenok:2014pua,Astashenok:2013vza,Arapoglu:2010rz,Panotopoulos:2021sbf,Lobato:2020fxt,Numajiri:2021nsc},
see also
\cite{Pani:2014jra,Staykov:2014mwa,Horbatsch:2015bua,Silva:2014fca,Doneva:2013qva,Xu:2020vbs,Salgado:1998sg,Shibata:2013pra,Arapoglu:2019mun,Ramazanoglu:2016kul,AltahaMotahar:2019ekm,Chew:2019lsa,Blazquez-Salcedo:2020ibb,Motahar:2017blm,Odintsov:2021qbq,Odintsov:2021nqa,Oikonomou:2021iid,Pretel:2022rwx,Pretel:2022plg,Cuzinatto:2016ehv,Oikonomou:2023dgu,Odintsov:2023ypt,Oikonomou:2023lnh,Brax:2017wcj,Akarsu:2018zxl,SavasArapoglu:2019eil,Pretel:2021kgl,Lin:2021ijx,Alam:2023grx,Murshid:2023xsw,Mota:2024kjb,Alwan:2024lng}
and theoretical astrophysics,
\cite{Altiparmak:2022bke,Bauswein:2020kor,Vretinaris:2019spn,Bauswein:2020aag,Bauswein:2017vtn,Most:2018hfd,Rezzolla:2017aly,Nathanail:2021tay,Koppel:2019pys,Raaijmakers:2021uju,Most:2020exl,Ecker:2022dlg,Jiang:2022tps}.
For our study we shall use several piecewise polytropic EoSs
\cite{Read:2008iy,Read:2009yp}, and specifically the SLy
\cite{Douchin:2001sv}, the AP3-AP4 \cite{Akmal:1997ft}, the WFF1
\cite{Wiringa:1988tp}, the ENG \cite{Engvik:1995gn}, the MPA1
\cite{Muther:1987xaa}, the MS1 and MS1b \cite{Mueller:1996pm} and
also the APR EoS \cite{Akmal:1998cf,Schneider:2019vdm} and with
regard to the latter, it is shown that the APR EoS reproduces the
variational calculations of \cite{Akmal:1998cf}, as was explained
in \cite{Schneider:2019vdm}. Let us note that in principle one can
add quark matter EoSs in the study, for example
\cite{Alford:2017qgh}, instead of purely hadronic which we chose
to study, but we did not extend the analysis to quark matter EoSs
for uniformity and simplicity, with no particular physical
reasoning behind our choice.

\begin{figure}[h!]
\centering
\includegraphics[width=30pc]{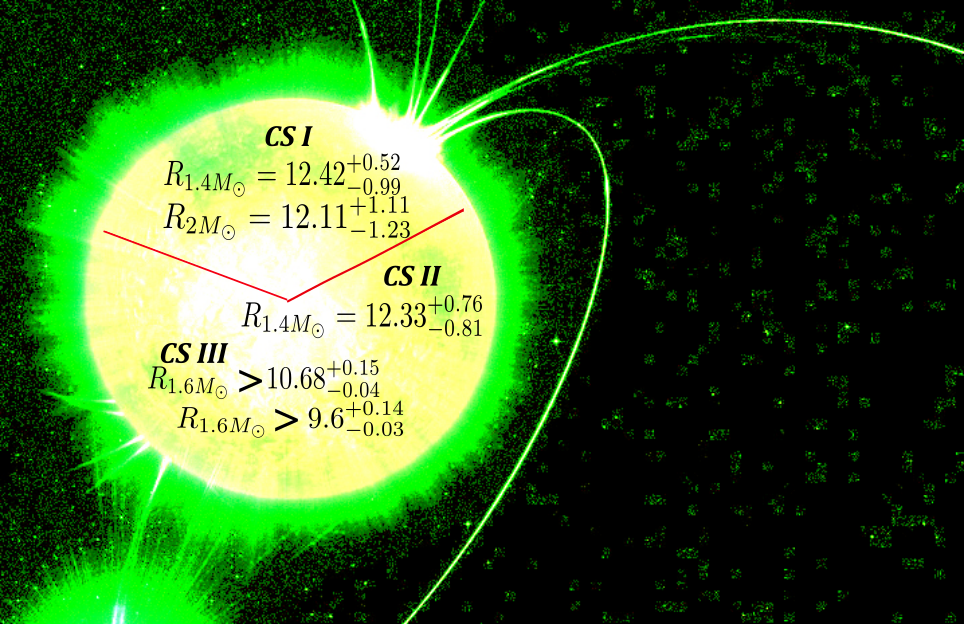}
\caption{The constraints CSI \cite{Altiparmak:2022bke}
$R_{1.4M_{\odot}}=12.42^{+0.52}_{-0.99}$ and
$R_{2M_{\odot}}=12.11^{+1.11}_{-1.23}\,$km, CSII
\cite{Raaijmakers:2021uju} with
$R_{1.4M_{\odot}}=12.33^{+0.76}_{-0.81}\,\mathrm{km}$ and CSIII
\cite{Bauswein:2017vtn} which indicates that the radius of a
$1.6M_{\odot}$ mass NS must satisfy
$R_{1.6M_{\odot}}>10.68^{+0.15}_{-0.04}\,$km and NSs with the
maximum mass, must have radius that satisfies
$R_{M_{max}}>9.6^{+0.14}_{-0.03}\,$km. This figure is based, after
heavy editing, on a public image of ESO, which can be found free
in Credit: ESO/L.Cal\c{c}ada:
\url{https://www.eso.org/public/images/eso0831a/.}} \label{plotcs}
\end{figure}

From previous studies for inflationary attractors
\cite{Odintsov:2023ypt}, the MPA1 seems to fit all the NS
phenomenological constraints. In the present work, the focus is on
supergravity motivated vector $f(R)$ gravity scalar-tensor
potentials \cite{Ozkan:2015iva}, which can be cosmologically
viable and non-viable. As we demonstrate, to our surprise even the
cosmologically non-viable vector $f(R)$ models produce a viable NS
phenomenology and the MPA1 is at the epicenter of the viable NS
phenomenologies. Technically, our numerical method to solve the
Tolman-Oppenheimer-Volkoff (TOV) equations is an LSODA python
based double-shooting method, that will yield the Jordan frame
Arnowitt-Deser-Misner (ADM) gravitational mass and radius of the
NS \cite{Arnowitt:1960zzc}. The NS phenomenological constraints we
shall use in order to test the vector $f(R)$ gravity models are
the NICER constraints, some recent modifications of NICER, the
constraints of PSR J0740+6620
\cite{Miller:2021qha,Providencia:2023rxc}, and also three
mainstream constraints which we shall refer to as CSI, CSII and
CSIII. The constraint CSI \cite{Altiparmak:2022bke} indicates that
the radius of a $1.4M_{\odot}$ mass NS
 has to be
$R_{1.4M_{\odot}}=12.42^{+0.52}_{-0.99}$ and the radius of an
$2M_{\odot}$ mass NS must be
$R_{2M_{\odot}}=12.11^{+1.11}_{-1.23}\,$km. The constraint CSII
\cite{Raaijmakers:2021uju} indicates that the radius of a
$1.4M_{\odot}$ mass NS has to be
$R_{1.4M_{\odot}}=12.33^{+0.76}_{-0.81}\,\mathrm{km}$, while the
constraint CSIII \cite{Bauswein:2017vtn} indicates that the radius
of an $1.6M_{\odot}$ mass NS must be larger than
$R_{1.6M_{\odot}}>10.68^{+0.15}_{-0.04}\,$km, and in addition, the
radius that corresponds to the maximum NS mass for a specified EoS
must be larger than $R_{M_{max}}>9.6^{+0.14}_{-0.03}\,$km. All the
phenomenological constraints CSI, CSII and CSIII are gathered for
convenience in Fig. \ref{plotcs}, and all the NS phenomenological
constraints appear in Table \ref{table0}.
\begin{table}[h!]
  \begin{center}
    \caption{\emph{\textbf{NS Phenomenological Constraints}}}
    \label{table0}
    \begin{tabular}{|r|r|}
     \hline
      \textbf{Constraint}   & \textbf{Mass and Radius}
      \\ \hline
      \textbf{CSI} & For $M=1.4M_{\odot}$, $R_{1.4M_{\odot}}=12.42^{+0.52}_{-0.99}$ and for $M=2M_{\odot}$, $R_{2M_{\odot}}=12.11^{+1.11}_{-1.23}\,$km.
\\  \hline
      \textbf{CSII} & For $M=1.4M_{\odot}$, $R_{1.4M_{\odot}}=12.33^{+0.76}_{-0.81}\,\mathrm{km}$.
\\  \hline
\textbf{CSIII} & For $M=1.6M_{\odot}$,
$R_{1.6M_{\odot}}>10.68^{+0.15}_{-0.04}\,$km, and for $M=M_{max}$,
$R_{M_{max}}>9.6^{+0.14}_{-0.03}\,$km.
\\  \hline
 NICER I & For $M=1.4M_{\odot}$, $11.34\,\mathrm{km}<R_{1.4M_{\odot}}<13.23\,\mathrm{km}$
\\  \hline
 NICER II & For $M=1.4M_{\odot}$,
 $12.33\,\mathrm{km}<R_{1.4M_{\odot}}<13.25\,\mathrm{km}$
\\  \hline
PSR J0740+6620& For $M=2.08M_{\odot}$,
 $11.6\,\mathrm{km}<R_{2.08M_{\odot}}<13.1\,\mathrm{km}$
\\  \hline
    \end{tabular}
  \end{center}
\end{table}
The result of our analysis indicates the importance of the
phenomenological EoS MPA1, which is greatly compatible with the
constraints for all the models of vector $f(R)$ gravity we used
for our analysis. What surprised us however, is the fact that even
cosmologically non-viable vector $f(R)$ gravity models yield a
viable NS phenomenology. Thus even non-viable vector $f(R)$
gravity inflationary models yield a viable NS phenomenology, with
the most refined scenario being related with the MPA1 EoS.

\section{Overview of the Scalar-tensor Formalism for Static Neutron Stars}

We shall briefly overview the formalism of Einstein frame
scalar-tensor theories and how the gravitational mass of the NS is
evaluated in these theories. Scalar-tensor theories in
astrophysical contexts are basically Einstein frame counterparts
of a known Jordan frame physical theory in the form of a
non-minimally coupled scalar field theory. Usually in
astrophysical contexts, geometrized units are used ($G=c=1$) and
also we shall use the notation of \cite{Pani:2014jra}. The Jordan
frame non-minimally coupled scalar field theory has the following
form,
\begin{equation}\label{taintro}
\mathcal{S}=\int
d^4x\frac{\sqrt{-g}}{16\pi}\Big{[}\Omega(\phi)R-\frac{1}{2}g^{\mu
\nu}\partial_{\mu}\phi\partial_{\nu}\phi-U(\phi)\Big{]}+S_m(\psi_m,g_{\mu
\nu})\, ,
\end{equation}
so after conformally transforming the above action, by using the
following transformation of the metric,
\begin{equation}\label{ta1higgsintro}
\tilde{g}_{\mu \nu}=A^{-2}g_{\mu \nu}\,
,\,\,\,A(\phi)=\Omega^{-1/2}(\phi)\, ,
\end{equation}
the Einstein frame action takes the following form,
\begin{equation}\label{ta5higgsintro}
\mathcal{S}=\int
d^4x\sqrt{-\tilde{g}}\Big{(}\frac{\tilde{R}}{16\pi}-\frac{1}{2}
\tilde{g}_{\mu \nu}\partial^{\mu}\varphi
\partial^{\nu}\varphi-\frac{V(\varphi)}{16\pi}\Big{)}+S_m(\psi_m,A^2(\varphi)\tilde{g}_{\mu
\nu})\, ,
\end{equation}
where $\varphi$ denotes the Einstein frame scalar field, with
$V(\varphi)$ which in turn is related to the Jordan frame scalar
field potential $U(\phi)$ as follows,
\begin{equation}\label{potentialns1intro}
V(\varphi)=\frac{U(\phi)}{\Omega^2}\, .
\end{equation}
There is an important function related to the conformal
transformation, that will also enter the TOV equations, the
function $\alpha(\varphi)$ defined in the following way,
\begin{equation}\label{alphaofvarphigeneraldefintro}
\alpha(\varphi)=\frac{d \ln A(\varphi)}{d \varphi}\, ,
\end{equation}
where  $A(\varphi)=\Omega^{-1/2}(\phi)$.
The metric that describes static NSs is the following,
\begin{equation}\label{tov1intro}
ds^2=-e^{\nu(r)}dt^2+\frac{dr^2}{1-\frac{2
m(r)}{r}}+r^2(d\theta^2+\sin^2\theta d\phi^2)\, ,
\end{equation}
where $m(r)$ is the mass function which describes the NS
gravitational mass, and $r$ denotes the circumferential radius.
Our numerical analysis that will follow aims in solving the TOV
equations and obtain numerically the metric function $\nu(r)$ and
the gravitational mass function $\frac{1}{1-\frac{2 m(r)}{r}}$. It
is important to stress that in modified gravity theories the
gravitational mass of the NS receives contribution beyond the
surface of the NS, in contrast with ordinary GR studies. Thus the
metric of the NS beyond the surface of the star is not directly a
Schwarzschild but it is a Schwarzschild one at the numerical
infinity. We shall discuss this important issue later on in this
and in the following sections. Proceeding to the analysis of
scalar-tensor NSs, if we assume that ordinary matter with pressure
$P$ and energy density $\epsilon$ is present, then by varying the
gravitational action we obtain the TOV equations,
\begin{equation}\label{tov2intro}
\frac{d m}{dr}=4\pi r^2
A^4(\varphi)\varepsilon+\frac{r}{2}(r-2m(r))\omega^2+4\pi
r^2V(\varphi)\, ,
\end{equation}
\begin{equation}\label{tov3intro}
\frac{d\nu}{dr}=r\omega^2+\frac{2}{r(r-2m(r))}\Big{[}4\pi
A^4(\varphi)r^3P-4\pi V(\varphi)
r^3\Big{]}+\frac{2m(r)}{r(r-2m(r))}\, ,
\end{equation}
\begin{equation}\label{tov4intro}
\frac{d\omega}{dr}=\frac{4\pi r
A^4(\varphi)}{r-2m(r)}\Big{(}\alpha(\varphi)(\epsilon-3P)+
r\omega(\epsilon-P)\Big{)}-\frac{2\omega
(r-m(r))}{r(r-2m(r))}+\frac{8\pi \omega r^2 V(\varphi)+r\frac{d
V(\varphi)}{d \varphi}}{r-2 m(r)}\, ,
\end{equation}
\begin{equation}\label{tov5intro}
\frac{dP}{dr}=-(\epsilon+P)\Big{[}\frac{1}{2}\frac{d\nu}{dr}+\alpha
(\varphi)\omega\Big{]}\, ,
\end{equation}
\begin{equation}\label{tov5newfinalintro}
\omega=\frac{d \varphi}{dr}\, ,
\end{equation}
where the function $\alpha (\varphi)$ was defined in Eq.
(\ref{alphaofvarphigeneraldefintro}). Now an important issue
related to the discussion regarding the gravitational mass
receiving contributions from beyond the star, due to the presence
of the scalar field, is the choice of the initial conditions,
which are the following,
\begin{equation}\label{tov8intro}
P(0)=P_c\, ,\,\,\,m(0)=0\, , \,\,\,\nu(0)=-\nu_c\, ,
\,\,\,\varphi(0)=\varphi_c\, ,\,\,\, \omega (0)=0\, .
\end{equation}
The choices for the metric function value $\nu_c$ and for the
scalar field value $\varphi_c$ at the center of the star, are
arbitrary, but the correct choice for them will be revealed by
using rigid optimization methods. We shall use a double shooting
method for obtaining the values of these parameters, which shall
be based on the fact that the values of the scalar field at
numerical infinity must be zero. Thus starting by arbitrary values
initially, the double shooting method will deliver to us the
correct values that make the scalar field vanish at numerical
infinity, at which point the metric is demanded to be a
Schwarzschild one. Now with regard to the matter that composes the
NS, we shall use a piecewise polytropic type of equation of state
\cite{Read:2008iy,Read:2009yp}, which in principle can be
generated for all the known EoSs. Specifically we shall use the
piecewise polytropic versions of the SLy \cite{Douchin:2001sv},
the WFF1 \cite{Wiringa:1988tp}, the AP3-AP4 \cite{Akmal:1998cf},
the ENG \cite{Engvik:1995gn}, the MPA1 \cite{Muther:1987xaa}, the
MS1 and MS1b \cite{Mueller:1996pm} and also the APR EoS
\cite{Akmal:1997ft}.

An important feature brought into play by modified gravity
theories is the fact that the NS receives contribution to its
gravitational mass beyond the surface of the star. This is due to
the modified gravity effects, either materialized by the scalar
field or the higher metric derivatives in $f(R)$ gravity. Thus it
is vital to extract a formula for the gravitational mass in
scalar-tensor theories. We shall calculate the ADM mass in the
Einstein frame for the static NS. To this end we introduce the
following quantities $K_E$ and $K_J$,
\begin{equation}\label{hE}
\mathcal{K}_E=1-\frac{2 m}{r_E}\, ,
\end{equation}
\begin{equation}\label{hE}
\mathcal{K}_J=1-\frac{2  m_J}{r_J}\, ,
\end{equation}
which are conformally related in the following way,
\begin{equation}\label{hehjrelation}
\mathcal{K}_J=A^{-2}\mathcal{K}_E\, .
\end{equation}
Also the radii of the NS in the Jordan and the Einstein frame are
connected as follows,
\begin{equation}\label{radiiconftrans}
r_J=A r_E\, .
\end{equation}
The Jordan frame ADM gravitational mass of the NS has the
following form,
\begin{equation}\label{jordaframemass1}
M_J=\lim_{r\to \infty}\frac{r_J}{2}\left(1-\mathcal{K}_J \right)
\, ,
\end{equation}
and the corresponding Einstein frame ADM gravitational mass has
the following form,
\begin{equation}\label{einsteiframemass1}
M_E=\lim_{r\to \infty}\frac{r_E}{2}\left(1-\mathcal{K}_E \right)
\, .
\end{equation}
Asymptotically from Eq. (\ref{hehjrelation}) we get,
\begin{equation}\label{asymptotich}
\mathcal{K}_J(r_E)=\left(1+\alpha(\varphi(r_E))\frac{d \varphi}{d
r}r_E \right)^2\mathcal{K}_E(\varphi(r_E))\, ,
\end{equation}
where $r_E$ stands for the Einstein frame radius parameter at
numerical infinity and furthermore $\frac{d\varphi
}{dr}=\frac{d\varphi }{dr}\Big{|}_{r=r_E}$. Upon combining Eqs.
(\ref{hE})-(\ref{asymptotich}) we obtain the following formula for
the Jordan frame ADM gravitational mass for the NS,
\begin{equation}\label{jordanframeADMmassfinal}
M_J=A(\varphi(r_E))\left(M_E-\frac{r_E^{2}}{2}\alpha
(\varphi(r_E))\frac{d\varphi
}{dr}\left(2+\alpha(\varphi(r_E))r_E\frac{d \varphi}{dr}
\right)\left(1-\frac{2 M_E}{r_E} \right) \right)\, ,
\end{equation}
with $\frac{d\varphi }{dr}=\frac{d\varphi }{dr}\Big{|}_{r=r_E}$.
\begin{figure}[h!]
\centering
\includegraphics[width=18pc]{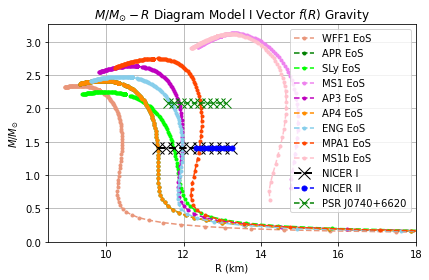}
\caption{The $M-R$ graphs for the Model I for the WFF1, SLy, APR,
MS1, AP3, AP4, ENG, MPA1, MS1b EoSs. We included the NICER I
\cite{Miller:2021qha}, NICER II \cite{Ecker:2022dlg} and the PSR
J0740+6620  constraints PSR J0740+6620
\cite{Miller:2021qha,Providencia:2023rxc}.} \label{plot1}
\end{figure}
In addition, the circumferential radius of the NS in the Jordan
frame, denoted as $R$, and the Einstein frame, denoted as $R_s$,
are related as follows,
\begin{equation}\label{radiussurface}
R=A(\varphi(R_s))\, R_s\, .
\end{equation}
With our numerical analysis, we shall extract the Jordan frame
masses and radii of NS in vector $f(R)$ gravity theories, by
firstly obtaining their Einstein frame counterparts. The
importance of the Jordan frame is profound, since in this frame,
matter follows free fall geodesics and matter is not coupled to
the metric. This is why any $M-R$ graph for NSs must contain only
Jordan frame quantities.
\begin{figure}[h!]
\centering
\includegraphics[width=18pc]{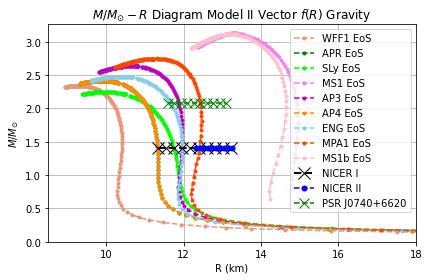}
\caption{The $M-R$ graphs for the Model II for the WFF1, SLy, APR,
MS1, AP3, AP4, ENG, MPA1, MS1b EoSs. We included the NICER I
\cite{Miller:2021qha}, NICER II \cite{Ecker:2022dlg} constraints
and PSR J0740+6620 constraints
\cite{Miller:2021qha,Providencia:2023rxc}.} \label{plot2}
\end{figure}

\subsection{Inflation and Neutron Stars Phenomenology with vector $f(R)$ Gravity}

Vector $f(R)$ gravity models of gravity \cite{Ozkan:2015iva} are
generated by replacing the Ricci scalar $R$ by
$R+A_{\mu}A^{\mu}+\beta\nabla_{\mu}A^{\mu}$, where $A_{\mu}$ is an
auxiliary vector field and $\beta$ is some positive parameter. The
resulting theory is basically equivalent to a Brans-Dicke theory
with Brans-Dicke parameter $\omega_{BD}=\frac{\beta^2}{4}$ with
only one scalar propagating degree of freedom. The whole framework
of auxiliary vector field enhanced $f(R)$ gravity is motivated by
supersymmetric extensions of the Starobinsky model
\cite{Kallosh:2013lkr,Ferrara:2013rsa,Kallosh:2013yoa}.
Specifically, in the old minimal $N=1$ off-shell supergravity, the
Weyl multiplet consists of the vielbein, the gravitino, an
auxiliary vector field $A_{\mu}$ and an auxiliary scalar field.
Embedding the $R^2$ model in this framework is done by coupling a
chiral multiplet to the Weyl multiple. Accordingly, the
supersymmetric version of the $R^2$ model can be cast in the form
of a scalar-tensor theory by simply integrating out the auxiliary
fields. We shall follow the model analysis and framework of Ref.
\cite{Ozkan:2015iva} in the following.
\begin{figure}[h!]
\centering
\includegraphics[width=18pc]{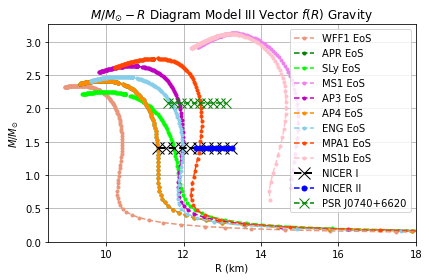}
\caption{The $M-R$ graphs for the Model III for the WFF1, SLy,
APR, MS1, AP3, AP4, ENG, MPA1, MS1b EoSs. We included the NICER I
\cite{Miller:2021qha}, NICER II \cite{Ecker:2022dlg} constraints
and PSR J0740+6620 constraints
\cite{Miller:2021qha,Providencia:2023rxc}.} \label{plot3}
\end{figure}
The vector $R^2$ model is described by the following Lagrangian
density,
\begin{equation}\label{vfr1}
\mathcal{L}=R+A_{\mu}A^{\mu}+\beta A_{\mu}\nabla^{\mu}+\frac{1}{6
M^2}\left(R+A_{\mu}A^{\mu}+\beta\nabla_{\mu}A^{\mu} \right)^2\, .
\end{equation}
Upon rewriting the Lagrangian as follows, by introducing an
auxiliary Lagrange multiplier scalar field $\phi$ and $F$,
\begin{equation}\label{vfr2}
\mathcal{L}=F+\frac{1}{6M^2}F^2-\phi\left(F-R-A_{\mu}A^{\mu}-\beta
A_{\mu}\nabla^{\mu} \right)\, .
\end{equation}
Upon varying the above with respect to the auxiliary fields
$A_{\mu}$ and $F$, we obtain the following equations,
\begin{equation}\label{efrv}
A_{\mu}=\frac{1}{2\phi}\beta
\nabla_{\mu}\phi,\,\,\,F=3M^2(\phi-1)\, .
\end{equation}
The equation above that involves the auxiliary vector field
indicates that on-shell, the vector field is equivalent to the
gradient of a scalar field. Combining the field equations and
integrating the action, by omitting a total derivative term, the
Lagrangian reads,
\begin{equation}\label{frv3}
\mathcal{L}=\phi R-\frac{1}{4\phi}\beta
\nabla_{\mu}\phi\nabla^{\mu}\phi-\frac{3}{2}M^2(\phi-1)^2\, ,
\end{equation}
so upon performing the conformal transformation of Eq.
(\ref{ta1higgsintro}), namely $\tilde{g}_{\mu \nu}=A^{-2}g_{\mu
\nu}$, we get the Einstein frame action, in the presence of
matter, and in Geometrized units.
\begin{figure}[h!]
\centering
\includegraphics[width=18pc]{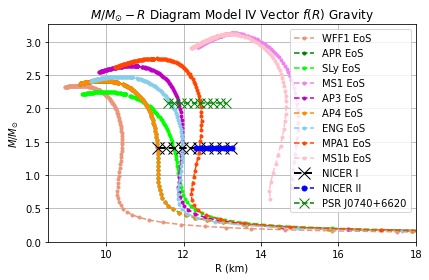}
\caption{The $M-R$ graphs for the Model IV for the WFF1, SLy, APR,
MS1, AP3, AP4, ENG, MPA1, MS1b EoSs. We included the NICER I
\cite{Miller:2021qha}, NICER II \cite{Ecker:2022dlg} constraints
and PSR J0740+6620 constraints
\cite{Miller:2021qha,Providencia:2023rxc}.} \label{plot4}
\end{figure}
\begin{equation}\label{frv4}
\mathcal{S}=\int
d^4x\sqrt{-\tilde{g}}\Big{(}\frac{\tilde{R}}{16\pi}-\frac{1}{2}
\tilde{g}_{\mu \nu}\partial^{\mu}\varphi
\partial^{\nu}\varphi-\frac{V(\varphi)}{16\pi}\Big{)}+S_m(\psi_m,A^2(\varphi)\tilde{g}_{\mu
\nu})\, ,
\end{equation}
where the potential $V(\varphi)$ for the Starobinsky model is,
\begin{equation}\label{starobpote}
V(\varphi)=\frac{3}{4}M^2\left(1-e^{-\sqrt{\frac{2}{3\alpha}}\varphi}
\right)^2\, ,
\end{equation}
with $\alpha$ being defined as,
\begin{equation}\label{alphasmalldefinition}
\alpha=1+\frac{\beta^2}{6}\, .
\end{equation}
The above formalism can be extended for generalized forms of
$f(R)$ gravity, so starting from a Lagrangian density,
\begin{equation}\label{frv5}
\mathcal{L}=f(R+A_{\mu}A^{\mu}+\beta A_{\mu}\nabla^{\mu})\, ,
\end{equation}
and upon rewriting it as,
\begin{equation}\label{frv6}
\mathcal{L}=f(F)-\phi(F-R-A_{\mu}A^{\mu}-\beta
A_{\mu}\nabla^{\mu})\, ,
\end{equation}
and varying with respect to $A_{\mu}$ and $F$, we get,
\begin{equation}\label{frv7}
A_{\mu}=\frac{1}{2\phi}\beta \nabla_{\mu}\phi,\,\,\,\frac{\partial
f}{\partial F}=\phi\, .
\end{equation}
So upon substituting (\ref{frv7}) in (\ref{frv6}) we get,
\begin{equation}\label{frv8}
\mathcal{L}=\phi R-\frac{1}{4\phi}\beta
\nabla_{\mu}\phi\nabla^{\mu}\phi-(\phi F(\phi)-f(F(\phi)))\, ,
\end{equation}
so the Jordan frame potential is $U(\phi)=\phi
F(\phi)-f(F(\phi))$. Upon performing the conformal transformation
$\tilde{g}_{\mu \nu}=\phi g_{\mu \nu}$, we get the Einstein frame
action, with the Einstein frame potential being,
\begin{equation}\label{frv9}
V(\varphi)=2^{-1}e^{-\sqrt{\frac{2}{3\alpha}}\varphi
}\left(F-e^{-\sqrt{\frac{2}{3\alpha}}\varphi }f(F)\right)\, ,
\end{equation}
and recall $\alpha$ is defined in Eq. (\ref{alphasmalldefinition})
and also $\frac{\partial f}{\partial
F}=\phi=e^{\sqrt{\frac{2}{3\alpha}}\varphi}$. Now let us choose
several models of vector $f(R)$ gravity for our NS study, and also
we shall make contact with the notation of the previous section
and specify all the functions and parameters that enter the TOV
equations.
\begin{figure}[h!]
\centering
\includegraphics[width=18pc]{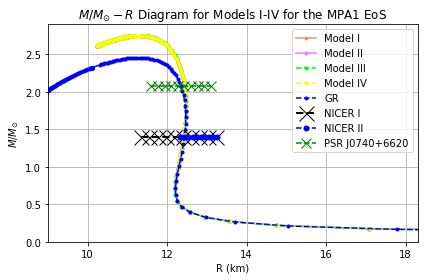}
\caption{The $M-R$ graphs for Models I-IV for the MPA1 EoS,
including the GR $M-R$ curve. An unexpected result is that the
models are almost indistinguishable, and this result holds true
for all the EoSs.} \label{plot5}
\end{figure}
For all the forthcoming scenarios, the Jordan frame scalar field
$\phi$ and the Einstein frame canonical scalar field are related
as follows,
\begin{equation}\label{furfg}
\phi=e^{\frac{2\varphi}{\sqrt{6+\beta^2}}}\, ,
\end{equation}
and the function $A(\varphi)$ related with the conformal
transformation, and defined in Eq. (\ref{ta1higgsintro}), as a
function of the canonical scalar field $\varphi$ reads,
\begin{equation}\label{af}
A(\varphi)=e^{\frac{\varphi}{2\sqrt{6+\beta^2}}}\, ,
\end{equation}
and also the function $\alpha(\varphi)$ defined in Eq.
(\ref{alphaofvarphigeneraldefintro}) reads,
\begin{equation}\label{afphi}
\alpha(\varphi)=\frac{1}{2\sqrt{6+\beta^2}}\, .
\end{equation}
Now let us define the models of vector $f(R)$ gravity which we
shall consider, and we shall focus on inflationary models. Firstly
we shall consider the $R^2$ model in which case the potential
reads,
\begin{equation}\label{starob1model}
V(\varphi)=\frac{3M^2}{4}\left(1-e^{-\sqrt{\frac{2}{3}\varphi}}
\right)\, ,
\end{equation}
so $\beta=0$ in this case, and also the viability of the
inflationary era, and specifically the constraints from the Planck
data on the amplitude of the scalar perturbations, indicate that
the parameter $M$ must be in this case, $M=1.3\times
10^{-5}\sqrt{1+\frac{\beta^2}{6}}\left(\frac{N}{55}\right)^{-1}$,
where $N$ is the $e$-foldings number which we shall take equal to
$N\sim 60$. We shall refer to this model as ``Model I'' hereafter.
Now a variant model of the Starobinsky model which we shall
consider is the model with potential,
\begin{equation}\label{starob1modelvar1}
V(\varphi)=\frac{3M^2}{4}\left(1-e^{-\sqrt{\frac{2}{3(1+\frac{\beta^2}{6})}\varphi}}
\right)\, ,
\end{equation}
with $\beta\sim 0.1$, which also yields a viable inflationary era,
and we shall refer to this model as ``Model II'' hereafter. Note
that we chose $\beta\sim 0.1$ because it is a value for which the
inflationary model of Eq. (\ref{starob1modelvar1}) yields
viability when confronted with the Planck 2018 constraints
\cite{Ozkan:2015iva}. Of course there are other values of $\beta$
close to $\beta\sim 0.1$ but we chose one for simplicity. This
class of models is characteristic and we dubbed them as Model II.
Also we shall consider another class of models originating from a
power-law $f(R)$ gravity, in which case the $f(R)$ gravity has the
form $f(R)=R+m^{2(1-n)}R^n$, where $n$ and $m$ will be constrained
by the viability of the inflationary era. The potential for this
theory in the Einstein frame reads,
\begin{equation}\label{potentialbigonepowerlaw}
V(\varphi)=\frac{n-1}{2n^{n/(n-1)}}m^2\,e^{-2\varphi\sqrt{\frac{2}{3(1+\frac{\beta^2}{6})}}}\left(e^{2\varphi\sqrt{\frac{2}{3(1+\frac{\beta^2}{6})}}}-1
\right)^{n/(n-1)}\, ,
\end{equation}
and the viability of the inflationary theory comes when $\beta=1$
and $1.75<n<2.39$. Also the parameter $m$ reads $m=5.1\times
10^{-4}pc_n^{-1/2}(2pN)^{-(p+2)/4}$, with
$c_n=(n-1)(2/(1+\frac{\beta^2}{6}))^{p/2}/2n^2$ and $p=n/(n-1)$.
We shall take $n=1.8$ for simplicity and we shall call this model
``Model III'' hereafter. Note that any value of $n$ in the range
$1.75<n<2.39$ is also correct, but we chose one characteristic
value for simplicity.  Finally we shall consider a limiting case
of this model, for $\beta\gg 1$ which we shall call Model IV, in
which case the potential reads,
\begin{equation}\label{finalpotentialread}
V(\varphi)=\frac{n-1}{2n^2}\left(
2/(1+\frac{\beta^2}{6})\right)^{p/2}m^2\varphi^p\, ,
\end{equation}
and we shall take in this case $n=4$ and $\beta=10^4$, which are
again characteristic values for this model. This model produces a
non-viable inflationary cosmology. In all the above cases, we used
geometrized units and in the following sections we shall solve
numerically the TOV equations and analyze in details the NS
phenomenology for each of the models I-IV.
\begin{table}[h!]
  \begin{center}
    \caption{\emph{\textbf{Maximum Masses for Vector $f(R)$ Gravity Models I-IV in the Mass Gap Region.}}}
    \label{tablemaxmasses}
    \begin{tabular}{|r|r|r|r|r|}
     \hline
      \textbf{Model}   & \textbf{MPA1 EoS} & \textbf{MS1b EoS} &
      \textbf{AP3 EoS} & \textbf{MS1 EoS}
      \\  \hline
      Model I & $M_{MPA1}= 2.7491013\,M_{\odot}$ & $M_{MS1b}= 3.1183\, M_{\odot}$ & $M_{AP3}= 2.63581\,
M_{\odot}$ & $M_{MS1}= 3.1269\,M_{\odot}$
\\  \hline
 Model II & $M_{MPA1}= 2.7491313\,M_{\odot}$ & $M_{MS1b}= 3.11788\, M_{\odot}$ & $M_{AP3}= 2.63619\,
M_{\odot}$ & $M_{MS1}= 3.1261\,M_{\odot}$
\\  \hline
 Model III & $M_{MPA1}= 2.749149\,M_{\odot}$ & $M_{MS1b}= 3.11837\, M_{\odot}$ & $M_{AP3}= 2.6359\,
M_{\odot}$ & $M_{MS1}= 3.12669\,M_{\odot}$
\\  \hline
 Model IV & $M_{MPA1}= 2.749149112\,M_{\odot}$ & $M_{MS1b}= 3.1178\, M_{\odot}$ & $M_{AP3}= 2.63618\,
M_{\odot}$ & $M_{MS1}= 3.12608\,M_{\odot}$
\\  \hline
    \end{tabular}
  \end{center}
\end{table}

\begin{table}[h!]
  \begin{center}
    \caption{\emph{\textbf{Vector $f(R)$ Gravity NSs vs CSI for NS Masses $M\sim 2M_{\odot}$, $R_{2M_{\odot}}=12.11^{+1.11}_{-1.23}\,$km, for the SLy, APR, WFF1, MS1 and AP3 EoSs. The ''x'' denotes non-viability.}}}
    \label{tablecsi2}
    \begin{tabular}{|r|r|r|r|r|r|}
     \hline
      \textbf{Model}   & \textbf{SLy EoS} & \textbf{APR EoS} & \textbf{WFF1
      EoS} & \textbf{MS1 EoS} & \textbf{AP3 EoS}
      \\  \hline
      \textbf{Model I} & $R_{SLy}= 11.15792\,$Km & $R_{APR}=  11.06273\,$Km & $R_{WFF1}=
      x$
       & $R_{MS1}= x$ & $R_{AP3}= 11.8980\,$Km
\\  \hline
\textbf{Model II} & $R_{SLy}= 11.14956\,$Km & $R_{APR}=
11.03405\,$Km & $R_{WFF1}=
      x$
       & $R_{MS1}= x$ & $R_{AP3}= 11.89015\,$Km
\\  \hline
\textbf{Model III} & $R_{SLy}= 11.15874\,$Km & $R_{APR}=
11.06346\,$Km & $R_{WFF1}=
      x$
       & $R_{MS1}= x$ & $R_{AP3}= 11.8989\,$Km
\\  \hline
\textbf{Model IV} & $R_{SLy}= 11.14957\,$Km & $R_{APR}=
11.08163\,$Km & $R_{WFF1}=
      x$
       & $R_{MS1}= x$ & $R_{AP3}= 11.89016\,$Km
\\  \hline
    \end{tabular}
  \end{center}
\end{table}

\begin{table}[h!]
  \begin{center}
    \caption{\emph{\textbf{Vector $f(R)$ Gravity NSs vs CSI for NS Masses $M\sim
2M_{\odot}$, $R_{2M_{\odot}}=12.11^{+1.11}_{-1.23}\,$km, for the
AP4, ENG, MPA1 and MS1b. The ''x'' denotes non-viability.}}}
    \label{tablecsi22}
    \begin{tabular}{|r|r|r|r|r|}
     \hline
      \textbf{Model}   & \textbf{AP4 EoS} & \textbf{ENG EoS} &
      \textbf{MPA1 EoS} & \textbf{MS1b EoS}
      \\  \hline
      \textbf{Model I} & $R_{AP4}= 11.650\,$Km & $R_{ENG}=  12.263\,$Km & $R_{MPA1}=
      13.014\,$Km
       & $R_{MS1b}= x$
\\  \hline
\textbf{Model II} & $R_{AP4}= 11.650\,$Km & $R_{ENG}= 12.263\,$Km
& $R_{MPA1}=
      13.014\,$Km
       & $R_{MS1b}= x$
\\  \hline
\textbf{Model III} & $R_{AP4}= 11.016096\,$Km & $R_{ENG}=
11.749539\,$Km & $R_{MPA1}=
      12.44922\,$Km
       & $R_{MS1b}= x$
\\  \hline
\textbf{Model IV} & $R_{AP4}= 11.081631\,$Km & $R_{ENG}=
11.74068\,$Km & $R_{MPA1}=
      12.44050\,$Km
       & $R_{MS1b}= x$
\\  \hline
    \end{tabular}
  \end{center}
\end{table}

\subsection{Results on the Phenomenology of NSs for the models I-IV and Viability of the Scenarios}

In this section we shall analyze the phenomenology of the models
I-IV developed in the previous section, by solving numerically the
TOV equations for each model presented. The numerical method we
shall adopt is based on a python LSODA solver, a variant of the
one developed in Ref. \cite{niksterg}. The method uses a double
shooting method to determine the optimal values of $\nu_c$ and
$\varphi_c$ at the center of the NS, which make the scalar field
vanish at numerical infinity. Special caution must be given in
determining the correct numerical infinity value for the radius
variable.
\begin{table}[h!]
  \begin{center}
    \caption{\emph{\textbf{Vector $f(R)$ Gravity NSs vs CSI for NS Masses $M\sim 1.4M_{\odot}$, $R_{1.4M_{\odot}}=12.42^{+0.52}_{-0.99}$, for the SLy, APR, WFF1, MS1 and AP3 EoSs. The ''x'' denotes non-viability. }}}
    \label{tablecsi14}
    \begin{tabular}{|r|r|r|r|r|r|}
     \hline
      \textbf{Model}   & \textbf{SLy EoS} & \textbf{APR EoS} & \textbf{WFF1
      EoS} & \textbf{MS1 EoS} & \textbf{AP3 EoS}
      \\  \hline
      \textbf{Model I} & $R_{SLy}= 11.73607\,$Km & $R_{APR}=  x$ & $R_{WFF1}=
      x$
       & $R_{MS1}= x$ & $R_{AP3}= 11.96694$
\\  \hline
 \textbf{Model II} & $R_{SLy}= 11.733879\,$Km & $R_{APR}= x$ & $R_{WFF1}=
      x$
       & $R_{MS1}= x$ & $R_{AP3}= 11.964894\,$Km
\\  \hline
 \textbf{Model III} & $R_{SLy}= 11.73665\,$Km & $R_{APR}=  x\,$Km & $R_{WFF1}=
      x$
       & $R_{MS1}= x$ & $R_{AP3}= 12.345$
\\  \hline
 \textbf{Model IV} & $R_{SLy}= 11.934\,$Km & $R_{APR}=  11.645\,$Km & $R_{WFF1}=
      x$
       & $R_{MS1}= x$ & $R_{AP3}= 12.333$
\\  \hline
    \end{tabular}
  \end{center}
\end{table}
One important thing to notice in the scalar-tensor studies of NSs
is that the gravitational mass of the NS receives contribution
beyond the surface of the star, due to the presence of the scalar
field. This can have drastic effects on the phenomenology of NS,
since the metric is not Schwarzschild outside the star but as the
radius tends to numerical infinity the metric becomes
asymptotically Schwarzschild. Our code will determine numerically
the Einstein frame masses and radii for NSs for the various EoSs
we mentioned in the introduction, and from these we shall evaluate
the corresponding Jordan frame quantities. Having the latter at
hand, we shall construct the $M-R$ graphs for all the EoSs and we
shall confront the resulting phenomenology with the NICER
constraints and with all the constraints appearing in Table
\ref{table0}. We quote the constraints CSI, CSII and CSIII here
for reading convenience, and CSI \cite{Altiparmak:2022bke}
constrains the radius of a NS with mass $1.4M_{\odot}$ and the
radius must be $R_{1.4M_{\odot}}=12.42^{+0.52}_{-0.99}$, and  for
the case of a $2M_{\odot}$ mass NS, the radius has to be
$R_{2M_{\odot}}=12.11^{+1.11}_{-1.23}\,$km. Also for CSII
\cite{Raaijmakers:2021uju}, a $1.4M_{\odot}$ mass NS, must have
radius $R_{1.4M_{\odot}}=12.33^{+0.76}_{-0.81}\,\mathrm{km}$.
Finally for CSIII, the radius of a $M=1.6M_{\odot}$ NS must be
$M=1.6M_{\odot}$, $R_{1.6M_{\odot}}>10.68^{+0.15}_{-0.04}\,$km,
while for the maximum mass of a NS, the radius must be larger than
$R_{M_{max}}>9.6^{+0.14}_{-0.03}\,$km.
\begin{table}[h!]
  \begin{center}
    \caption{\emph{\textbf{Vector $f(R)$ Gravity NSs vs CSI for NS Masses $M\sim
1.4M_{\odot}$, $R_{1.4M_{\odot}}=12.42^{+0.52}_{-0.99}$, for the
AP4, ENG, MPA1 and MS1b. The ''x'' denotes non-viability.}}}
    \label{tablecsi142}
    \begin{tabular}{|r|r|r|r|r|}
     \hline
      \textbf{Model}   & \textbf{AP4 EoS} & \textbf{ENG EoS} &
      \textbf{MPA1
      EoS} & \textbf{MS1b EoS}
      \\  \hline
      \textbf{Model I} & $R_{AP4}= x$ & $R_{ENG}=  11.973665\,$Km & $R_{MPA1}=
      12.415368\,$Km
       & $R_{MS1b}= x$
\\  \hline
 \textbf{Model II} & $R_{AP4}= x$ & $R_{ENG}=  11.97168\,$Km & $R_{MPA1}=
      12.41331\,$Km
       & $R_{MS1b}= x$
\\  \hline
 \textbf{Model III} & $R_{AP4}= x$ & $R_{ENG}=  11.974472\,$Km & $R_{MPA1}=
      12.415987\,$Km
       & $R_{MS1b}= x$
\\  \hline
 \textbf{Model IV} & $R_{AP4}= x\,$ & $R_{ENG}=  11.9716949\,$Km & $R_{MPA1}=
      12.413329\,$Km
       & $R_{MS1b}= x$
\\  \hline
    \end{tabular}
  \end{center}
\end{table}
In this section we shall present in detail all the
 NS phenomenological implications of Models I-IV.
\begin{table}[h!]
  \begin{center}
    \caption{\emph{\textbf{Vector $f(R)$ Gravity NSs Radii vs CSII for NS Masses $M\sim 1.4M_{\odot}$, $R_{1.4M_{\odot}}=12.33^{+0.76}_{-0.81}\,\mathrm{km}$, for the SLy, APR, WFF1, MS1 and AP3 EoSs. The ''x'' denotes non-viability.}}}
    \label{tablecsii14}
    \begin{tabular}{|r|r|r|r|r|r|}
     \hline
      \textbf{Model}   & \textbf{SLy EoS} & \textbf{APR EoS} & \textbf{WFF1
      EoS} & \textbf{MS1 EoS} & \textbf{AP3 EoS}
      \\  \hline
      \textbf{Model I} & $R_{SLy}= 11.73607\,$Km & $R_{APR}=  x$ & $R_{WFF1}=
      x$
       & $R_{MS1}= x$ & $R_{AP3}= 11.966948\,$Km
\\  \hline
\textbf{Model II} & $R_{SLy}= 11.73387\,$Km & $R_{APR}= x\,$ &
$R_{WFF1}=
      x$
       & $R_{MS1}= x$ & $R_{AP3}= 11.964894\,$Km
\\  \hline
\textbf{Model III} & $R_{SLy}= 11.7366\,$Km & $R_{APR}= x\,$ &
$R_{WFF1}=
      x$
       & $R_{MS1}= x$ & $R_{AP3}= 11.967712\,$Km
\\  \hline
\textbf{Model IV} & $R_{SLy}= 11.733891\,$Km & $R_{APR}= x\,$ &
$R_{WFF1}=
      x$
       & $R_{MS1}= x$ & $R_{AP3}= 11.96490\,$Km
\\  \hline
    \end{tabular}
  \end{center}
\end{table}
\begin{table}[h!]
  \begin{center}
    \caption{\emph{\textbf{Vector $f(R)$ Gravity NSs vs CSII for NS Masses $M\sim
1.4M_{\odot}$,
$R_{1.4M_{\odot}}=12.33^{+0.76}_{-0.81}\,\mathrm{km}$, for the
AP4, ENG, MPA1 and MS1b. The ''x'' denotes non-viability.}}}
    \label{tablecsii142}
    \begin{tabular}{|r|r|r|r|r|}
     \hline
      \textbf{Model}   & \textbf{AP4 EoS} & \textbf{ENG EoS} &
      \textbf{MPA1 EoS} & \textbf{MS1b EoS}\\  \hline
      \textbf{Model I} & $R_{AP4}= x\,$ & $R_{ENG}=  11.973665\,$Km & $R_{MPA1}=
      12.415368\,$Km  & $R_{MS1b}= x$
\\  \hline
\textbf{Model II} & $R_{AP4}= x\,$ & $R_{ENG}= 11.971680\,$Km &
$R_{MPA1}=
      12.413314\,$Km  & $R_{MS1b}= x$
\\  \hline
\textbf{Model III} & $R_{AP4}= x\,$ & $R_{ENG}= 1.437837\,$Km &
$R_{MPA1}=
      12.415987\,$Km  & $R_{MS1b}= x$
\\  \hline
\textbf{Model IV} & $R_{AP4}= x\,$ & $R_{ENG}= 11.97169\,$Km &
$R_{MPA1}=
      12.41332\,$Km  & $R_{MS1b}= x$
\\  \hline
    \end{tabular}
  \end{center}
\end{table}
We shall start our presentation with the $M-R$ graphs for models
I-IV using all the distinct EoSs we mentioned in the introduction.
In each $M-R$ graph, we shall also consider the NICER constraints,
which recall that $R_{1.4M_{\odot}}=11.34-13.23\,$km when a
$M=1.4M_{\odot}$ NS is considered \cite{Miller:2021qha}. Also we
shall consider a refinement of NICER, developed in
\cite{Ecker:2022dlg} which also takes into account the black-widow
binary pulsar PSR J0952-0607 which has mass $M=2.35\pm 0.17$
\cite{Romani:2022jhd} and we refer to this constraint, as NICER II
constraint. In addition, we shall consider the constraints from
the PSR J0740+6620 \cite{Miller:2021qha,Providencia:2023rxc}. In
Figs. \ref{plot1}-\ref{plot4} we present the $M-R$ graphs the
models I-IV of vector $f(R)$ gravity, confronted with the NICER I
and II constraints, the PSR J0740+6620 constraints
\cite{Miller:2021qha,Providencia:2023rxc} and also for all the
EoSs we mentioned in the introduction, that is for the WFF1, SLy,
APR, MS1, AP3, AP4, ENG, MPA1, MS1b. From Figs. \ref{plot1} it is
obvious that the MPA1 EoS plays an important role since it is
fully compatible with all the NICER constraints, while the AP3,
AP4, SLy and ENG EoSs are compatible with only the NICER I
constraint. The importance of the MPA1 EoS was also pointed out in
other similar works where inflationary and dark matter scalar
potentials were used, see for example
\cite{Odintsov:2023ypt,Oikonomou:2024yzj}. Also, it is almost
clear that the four models I-IV of vector $f(R)$ gravity produce
quite similar phenomenology. These are almost indistinguishable as
it can be seen in Fig. \ref{plot5}. Also the models deviate from
the GR result, as it can be seen in Fig. \ref{plot5}. The
indistinguishability feature is quite surprising and we did not
expected this, since we expected that the non-viable models of
inflation would lead to non-viable NS phenomenology based for
example on previous cases, like the Higgs model
\cite{Oikonomou:2021gzv}. It seems that this feature is somewhat
model dependent, and also it strongly depends on the form of the
functions $A(\varphi)$ and $\alpha (\varphi)$. Still, we did not
expect this intriguing result. Some small differences can be found
between models when one considers the maximum mass of NSs and the
constraints CSI-CSIII for models I-IV, as we now show.
\begin{table}[h!]
  \begin{center}
    \caption{\emph{\textbf{Vector $f(R)$ Gravity NSs vs CSIII for NS Masses $M\sim 1.6M_{\odot}$, $R_{1.6M_{\odot}}>10.68^{+0.15}_{-0.04}\,$km, for the SLy, APR, WFF1, MS1 and AP3 EoSs. The ''x'' denotes non-viability.}}}
    \label{tablecsiii16}
    \begin{tabular}{|r|r|r|r|r|r|}
     \hline
      \textbf{Model}   & \textbf{SLy EoS} & \textbf{APR EoS} & \textbf{WFF1EoS} & \textbf{MS1 EoS} & \textbf{AP3 EoS}
      \\  \hline
      \textbf{Model I} & $R_{SLy}= 11.62696\,$Km & $R_{APR}=  11.29406\,$Km & $R_{WFF1}=
     x\,$
       & $R_{MS1}= x$ & $R_{AP3}= x$
\\  \hline
\textbf{Model II} & $R_{SLy}= 11.645561\,$Km & $R_{APR}=
11.285455\,$Km & $R_{WFF1}=
     x\,$
       & $R_{MS1}= x$ & $R_{AP3}= x$
\\  \hline
\textbf{Model III} & $R_{SLy}= 11.648040\,$Km & $R_{APR}=
11.28795\,$Km & $R_{WFF1}=
     x\,$
       & $R_{MS1}= x$ & $R_{AP3}= x$
\\  \hline
\textbf{Model IV} & $R_{SLy}= 11.645564\,$Km & $R_{APR}=
11.285462\,$Km & $R_{WFF1}=
     x\,$
       & $R_{MS1}= x$ & $R_{AP3}= x$
\\  \hline
    \end{tabular}
  \end{center}
\end{table}
\begin{table}[h!]
  \begin{center}
    \caption{\emph{\textbf{Vector $f(R)$ Gravity NSs vs CSIII for NS Masses $M\sim
1.6M_{\odot}$, $R_{1.6M_{\odot}}>10.68^{+0.15}_{-0.04}\,$km, for
the AP4, ENG, MPA1 and MS1b. The ''x'' denotes non-viability.}}}
    \label{tablecsiii162}
    \begin{tabular}{|r|r|r|r|r|}
     \hline
      \textbf{Model}   & \textbf{AP4 EoS} & \textbf{ENG EoS} &
      \textbf{MPA1
      EoS} & \textbf{MS1b EoS}
      \\  \hline
      \textbf{Model I} & $R_{AP4}= 11.294067\,$Km & $R_{ENG}=  11.952955\,$Km & $R_{MPA1}=
      12.448903\,$Km
       & $R_{MS1b}= x\,$
\\  \hline
\textbf{Model II} & $R_{AP4}= 11.285455\,$Km & $R_{ENG}=
11.951299\,$Km & $R_{MPA1}=
      12.447289\,$Km
       & $R_{MS1b}= 14.553546\,$Km
\\  \hline
\textbf{Model III} & $R_{AP4}= 11.287954\,$Km & $R_{ENG}=
11.953813\,$Km & $R_{MPA1}=
      12.4496248\,$Km
       & $R_{MS1b}= x\,$
\\  \hline
\textbf{Model IV} & $R_{AP4}= 11.285462\,$Km & $R_{ENG}=
11.951311\,$Km & $R_{MPA1}=
      12.44730\,$Km
       & $R_{MS1b}= 14.55355\,$Km
\\  \hline
    \end{tabular}
  \end{center}
\end{table}
With the numerical analysis we obtained the data which we gathered
in several tables in the text. Specifically in Table
\ref{tablemaxmasses} we quote the maximum NS masses which belong
in the mass gap region and the corresponding EoSs which achieve
this along with the model. A notable feature is that all the
predicted masses are below the 3 solar masses causal limit, and
the EoSs which predict a maximum mass beyond that upper limit, are
proven to provide a non-viable NS phenomenology, as we demonstrate
shortly. In Tables \ref{tablecsi2}-\ref{tablecsi22} the vector
$f(R)$ gravity models are confronted with the CSI constraint,
considering 2 solar masses NSs, and also in Tables
\ref{tablecsi14}-\ref{tablecsi142} the NS phenomenology is
confronted with CSI when $M\sim 1.4M_{\odot}$ NSs are considered.
In addition, in Tables \ref{tablecsii14}-\ref{tablecsii142} the
vector $f(R)$ phenomenology is confronted with the constrain CSII,
and the same procedure for CSIII is presented in Tables
\ref{tablecsiiimax}-\ref{tablecsiiimax2}.
\begin{table}[h!]
  \begin{center}
    \caption{\emph{\textbf{Vector $f(R)$ Gravity NSs Maximum Masses and the Corresponding Radii vs CSIII, $R_{M_{max}}>9.6^{+0.14}_{-0.03}\,$km, for the SLy, APR, WFF1, MS1 and AP3 EoSs. The ''x'' denotes non-viability. }}}
    \label{tablecsiiimax}
    \begin{tabular}{|r|r|r|r|r|r|}
     \hline
      \textbf{Model}   & \textbf{APR EoS} & \textbf{SLy EoS} & \textbf{WFF1
      EoS} & \textbf{MS1 EoS} & \textbf{AP3 EoS}
      \\  \hline
      \textbf{Model I $M_{max}$} & $M_{APR}= 2.417\,M_{\odot}$ & $M_{SLy}= 2.248\, M_{\odot}$ & $M_{WFF1}= 2.341\,
M_{\odot}$ & $M_{MS1}= 3.126\,M_{\odot}$ & $M_{AP3}=
2.524\,M_{\odot}$
\\  \hline
\textbf{Model I Radii} & $R_{APR}= 9.897\,$Km & $R_{SLy}=
9.984\,$Km & $R_{WFF1}=
     9.293\,$Km
       & $R_{MS1}= 13.312\,$Km & $R_{AP3}= 11.375\,$Km
\\  \hline
\textbf{Model II $M_{max}$} & $M_{APR}= 2.192\,M_{\odot}$ &
$M_{SLy}= 10.866\, M_{\odot}$ & $M_{WFF1}= 2.342\, M_{\odot}$ &
$M_{MS1}= 3.126\,M_{\odot}$ & $M_{AP3}= 2.524\,M_{\odot}$
\\  \hline
\textbf{Model II Radii} & $R_{APR}= 10.866\,$Km & $R_{SLy}=
9.987\,$Km & $R_{WFF1}=
     9.308\,$Km
       & $R_{MS1}= 13.910\,$Km & $R_{AP3}= 11.369\,$Km
\\  \hline
\textbf{Model III $M_{max}$} & $M_{APR}= 2.417\,M_{\odot}$ &
$M_{SLy}= 2.248\, M_{\odot}$ & $M_{WFF1}= 2.342\, M_{\odot}$ &
$M_{MS1}= 3.126\,M_{\odot}$ & $M_{AP3}= 2.635\,M_{\odot}$
\\  \hline
\textbf{Model III Radii} & $R_{APR}= 9.917\,$Km & $R_{SLy}=
9.984\,$Km & $R_{WFF1}=9.293
     \,$Km
       & $R_{MS1}= 13.313\,$Km & $R_{AP3}= 10.651\,$Km
\\  \hline
\textbf{Model IV $M_{max}$} & $M_{APR}= 2.417\,M_{\odot}$ &
$M_{SLy}= 2.248\, M_{\odot}$ & $M_{WFF1}= 2.342\, M_{\odot}$ &
$M_{MS1}= 3.126\,M_{\odot}$ & $M_{AP3}= 2.636\,M_{\odot}$
\\  \hline
\textbf{Model IV Radii} & $R_{APR}= 9.899\,$Km & $R_{SLy}=
9.967\,$Km & $R_{WFF1}=
     9.281\,$Km
       & $R_{MS1}= 13.310\,$Km & $R_{AP3}= 10.673\,$Km
\\  \hline
    \end{tabular}
  \end{center}
\end{table}
\begin{table}[h!]
  \begin{center}
    \caption{\emph{\textbf{Vector $f(R)$ Gravity NSs Maximum Masses and the and the correspondent vs CSIII, $R_{M_{max}}>9.6^{+0.14}_{-0.03}\,$km, for the AP4, ENG, MPA1 and MS1b. The ''x'' denotes non-viability. }}}
    \label{tablecsiiimax2}
    \begin{tabular}{|r|r|r|r|r|}
     \hline
      \textbf{Model}   & \textbf{AP4 EoS} & \textbf{ENG EoS} &
      \textbf{MPA1
      EoS} & \textbf{MS1b EoS}
      \\  \hline
      \textbf{Model I $M_{max}$} & $M_{AP4}= 2.417\,M_{\odot}$ & $M_{ENG}= 2.478\, M_{\odot}$ & $M_{MPA1}= 2.749\,
M_{\odot}$ & $M_{MS1b}= 3.118\,M_{\odot}$
\\  \hline
\textbf{Model I Radii} & $R_{AP4}= 9.897\,$Km & $R_{ENG}=
10.385\,$Km & $R_{MPA1}=
      11.329\,$Km
       & $R_{MS1b}= 13.224\,$Km
\\  \hline
 \textbf{Model II $M_{max}$} & $M_{AP4}= 2.417\,M_{\odot}$ & $M_{ENG}=2.478\, M_{\odot}$ & $M_{MPA1}= 2.749\,
M_{\odot}$ & $M_{MS1b}= 3.117\,M_{\odot}$
\\  \hline
\textbf{Model II Radii} & $R_{AP4}= 9.912\,$Km & $R_{ENG}=
10.361\,$Km & $R_{MPA1}=
      11.326\,$Km
       & $R_{MS1b}= 13.215\,$Km
\\  \hline
 \textbf{Model III $M_{max}$} & $M_{AP4}= 2.417\,M_{\odot}$ & $M_{ENG}= 2.478\, M_{\odot}$ & $M_{MPA1}= 2.749\,
M_{\odot}$ & $M_{MS1b}= 3.118\,M_{\odot}$
\\  \hline
\textbf{Model III Radii} & $R_{AP4}= 9.917\,$Km & $R_{ENG}=
10.379\,$Km & $R_{MPA1}=
      11.330\,$Km
       & $R_{MS1b}= 13.238\,$Km
\\  \hline
 \textbf{Model IV $M_{max}$} & $M_{AP4}= 2.417\,M_{\odot}$ & $M_{ENG}= 2.478\, M_{\odot}$ & $M_{MPA1}= 2.749\,
M_{\odot}$ & $M_{MS1b}= 3.117\,M_{\odot}$
\\  \hline
\textbf{Model IV Radii} & $R_{AP4}= 9.899\,$Km & $R_{ENG}=
10.361\,$Km & $R_{MPA1}=
      11.326\,$Km
       & $R_{MS1b}= 13.215\,$Km
\\  \hline
    \end{tabular}
  \end{center}
\end{table}
From the all the tables containing the extracted data from the
numerical analysis, it is apparent that three equations of state
are entirely excluded, namely the WFF1, the MS1 and the MS1b EoSs.
Among all EoS, AP3, AP4, SLy, ENG, and MPA1 are mostly compatible
with all the NICER I constraint, but the MPA1 EoS enjoys an
elevated role since it is compatible with the NICER I and NICER II
constraints and the PSR J0740+6620 constraints
\cite{Miller:2021qha,Providencia:2023rxc}, but it also is
compatible with all the constraints CSI, CSII and CSIII, for all
the models I-IV. Thus one fundamental question is whether this
MPA1 EoS plays an important role in nature. This question can be
answered once new data from NS mergers are provided, and these
mergers must have components in the mass-gap region. In order to
pinpoint such mergers, we have to be lucky, since two things must
synergistically apply to succeed in catching such mergers, a
kilonova and mass components in the mass-gap region. We hope that
the future observations will provide evidence of such events.

\section*{Concluding Remarks}

In this work we studied the static NS phenomenology for a vector
$f(R)$ gravity theory. These theories in the Jordan frame contain
vector fields which are motivated by supergravity extensions of
the Starobinsky model. In the Einstein frame these theories can be
recast in a scalar-tensor form and we considered several
interesting models which can generate a viable inflationary era,
but we also considered some cosmologically no-viable models. The
initial question we had in mind is whether cosmologically
non-viable models can provide a viable NS phenomenology. The
answer was, to our surprise, that even cosmologically non-viable
models generate a viable NS phenomenology. This feature has to be
model dependent though, since in other cases, cosmologically
non-viable models provide a non-viable NS phenomenology. Regarding
the approach used for extracting the NS phenomenology, we
constructed the TOV equations for this vector $f(R)$ gravity
theory, and we used a double shooting method to extract the
correct initial conditions for the scalar field and the metric
function at the center of the NS, which generate the most refined
solution for the scalar field at the numerical infinity. The
characteristic of NS theories in the context of the scalar-tensor
theories is that the gravitational mass of the NS receives
contributions beyond the surface of the NS, due to the presence of
the scalar field. We used an LSODA python based code in order to
calculate the Einstein frame mass and radius of the NS, and from
these we calculated the corresponding Jordan frame quantities.
Regarding the matter fluid, we considered several
phenomenologically important EoSs, and specifically we considered
the WFF1, the SLy, the APR, the MS1, the AP3, the AP4, the ENG,
the MPA1 and the MS1b, in the context of a piecewise approach.
Using the numerical data we constructed the Jordan frame $M-R$
graphs, and we confronted the various models phenomenology with
several existing phenomenological constraints, like the NICER
constraint and one variant form of it \cite{Ecker:2022dlg} which
we dubbed NICER II, the PSR J0740+6620 constraints
\cite{Miller:2021qha,Providencia:2023rxc} and also several other
phenomenological constraints which we called CSI, CSII and CSIII
appearing in Table \ref{table0}. We considered four distinct
models, with variant cosmological importance, and the resulting
phenomenology indicates that among all the various EoSs, the MPA1
EoS enjoys an elevated role, since the results related to this EoS
are compatible with all the constraints we used. Interestingly
enough, the MPA1 vector $f(R)$ gravity models predict a maximum
mass for the NSs which is inside the mass-gap region, but below
the 3 solar masses limit known as causal limit. Now the question
is why the MPA1 EoS enjoys such elevated role among the various
distinct EoSs, does it play a fundamental role in nature?
Intriguingly the predictions of this EoS for scalar-tensor
theories is that NSs are allowed to have masses within the
mass-gap region. The answer to this question is not
straightforward, since observations of heavy NSs in the mass-gap
region are needed. There exist observations of massive components
in mergers with mass in the mass-gap region, but currently their
identity is unknown, so we anticipate the future observations to
shed light on this aspect of NS phenomenology. Also it is
important to include studies on the predictions of theories of
modified gravity for the tidal deformability, the moment of
inertia, the oscillation spectrum and so on. However in the
context of scalar-tensor gravity, these studies are technically
demanding, so we hope to address some of these issues in the
future.

We need to point out that the present theoretical context did not
reveal any new physics or curious predictions regarding NSs in
modified gravity. It just complied with the general behavior of
viable modified gravity models and also it respects the 3 solar
masses rule even for viable modified gravity models, see for
example the similar in spirit \cite{Oikonomou:2024yzj}.

\section*{Acknowledgments}

This research is funded by the Committee of Science of the
Ministry of Education and Science of the Republic of Kazakhstan
(Grant No. AP26194585) (V.K. Oikonomou).

\end{document}